\documentclass[english,aps,manuscript]{revtex4}
\usepackage[T1]{fontenc}
\usepackage[latin9]{inputenc}
\usepackage{graphicx}
\usepackage{esint}

\makeatletter
\@ifundefined{textcolor}{}
{%
 \definecolor{BLACK}{gray}{0}
 \definecolor{WHITE}{gray}{1}
 \definecolor{RED}{rgb}{1,0,0}
 \definecolor{GREEN}{rgb}{0,1,0}
 \definecolor{BLUE}{rgb}{0,0,1}
 \definecolor{CYAN}{cmyk}{1,0,0,0}
 \definecolor{MAGENTA}{cmyk}{0,1,0,0}
 \definecolor{YELLOW}{cmyk}{0,0,1,0}
 }

\makeatother

\usepackage{babel}

\begin{document}

\title{Test of a Jastrow-type wavefunction for a trapped few-body system
in one dimension}

\author{J. C. Cremon}

\address{Mathematical Physics, Lund University, SE-22100 Lund, Sweden}
\begin{abstract}
For a system with interacting quantum mechanical particles in a one-dimensional
harmonic oscillator, a trial wavefunction with simple structure based
on the solution of the corresponding two-particle system is suggested
and tested numerically. With the inclusion of a scaling parameter
for the distance between particles, at least for the very small systems
tested here the ansatz gives a very good estimate of the ground state
energy, with the error being of the order of $\sim1\%$ of the gap
to the first excited state.
\end{abstract}
\maketitle
For a quantum many-body system with interacting particles, a trial
wavefunction based on a product of functions for the motion of each
individual pair was suggested some time ago by Jastrow \cite{jastrow}
(who also referred to works by Mott, and Dingle \cite{dingle}). This
type of wavefunction forms the basis for many successful quantum Monte-Carlo
methods (see e.g. Ref. \cite{qmc-review}). In this study, a suggestion
in Ref. \cite{jastrow} -- to approximate the pair-motion function
with the solution from a system with only two particles -- is tested
for a specific system, a one-dimensional harmonic oscillator with
a few interacting quantum mechanical particles. One-dimensional quantum
systems have been studied immensely, see e.g. Refs. \cite{gogolin-book,giamarchi-book,rigol-review}
and references therein, though a majority of this research has focused
on infinite systems. For confined systems with limited number of particles,
various numerical approaches have been developed and used, see e.g.
Refs. . The understanding of one-dimensional systems has become increasingly
important with experimental developments such as ultra-cold atoms
trapped in quasi-1D confinements \cite{bloch-dalibard-zwerger-2008},
and electrons in carbon nanotubes \cite{charlier-blase-roche-RMP-2007}
or semiconductor quantum wires \cite{samuelson-et-al-2004}. For example,
Ref. \cite{jochim-a} reports about recent experiments on 1--10 ultracold
atoms with tunable interactions, in a quasi-1D harmonic confinement.

The Hamiltonian considered here, expressed in dimensionless units,
is\begin{equation}
H=\sum_{i=1}^{N}\biggl(-\frac{1}{2}\frac{d^{2}}{dx_{i}^{2}}+\frac{1}{2}x_{i}^{2}\biggl)+\sum_{i<j}^{N}V(x_{i}-x_{j})\label{eq:hamiltonian-N}\end{equation}
with the interaction $V$ depending only on the relative distance
between particles. For the ground state wave function, the ansatz
considered here is of the form

\begin{equation}
\psi(x_{1},...,x_{N})\propto\biggl[\prod_{k=1}^{N}e^{-x_{k}^{2}/2}\biggl]\biggl[\prod_{i<j}^{N}f\biggl(\frac{x_{i}-x_{j}}{\sqrt{2}}\biggl)\biggl]\label{eq:trial-wavefunction-N}\end{equation}
which can be shown to be exactly valid for a system with only two
particles, in which case a coordinate transformation separates the
Hamiltonian. As mentioned above, this type of wavefunction, based
on the relative motion of pairs of particles rather than single-particle
orbitals, was tried by Jastrow some time ago \cite{jastrow}. (Also
note the similarity with equation (6) in Ref. \cite{deuretzbacher}).
For a system with only one spatial dimension one can ensure bosonic
symmetry (or fermionic anti-symmetry) by choosing $f(x)$ as an even
(or odd) function. For example, a system with non-interacting bosons
(all in the lowest orbital) is then described with $f(x)=1$. For
fermions, setting $f(x)=x$ results in a Slater determinant with fermions
in the $N$ first orbitals (at least for small particle numbers this
is straightforward to check). This also implies that for bosons with
an infinitely strong repulsive zero-range interaction, a so-called
Tonks-Girardeau gas, the ground state is given by $f(x)=\vert x\vert$
\cite{tonks,girardeau}.

In this paper, the choice \[
f(x)=e^{+x^{2}/2}\phi(\alpha x)\]
is tested, where $\phi(x)$ is the relative wavefunction for the ground
state of a system with just two particles, and $\alpha$ is a scaling
factor. Such a scaling turns out to be necessary to get a reasonable
energy. To be more precise, $\phi(x)$ is the solution corresponding
to the lowest eigenvalue in the equation\[
\biggl(-\frac{1}{2}\frac{d^{2}}{dx^{2}}+\frac{x^{2}}{2}+V(\sqrt{2}\cdot x)\biggl)\phi(x)=E\phi(x)\]
where the factor of $\sqrt{2}$ is due to the coordinate transformation
$x=(x_{1}-x_{2})/\sqrt{2}$ that is used to transform to the relative
coordinate. For a general interaction $V$, $\phi(x)$ must be obtained
by numerical means. This is here done with a diagonalization approach,
with $\phi(x)$ expanded in harmonic oscillator orbitals (in the present
study, 10--15 orbitals are used). The energy of the many-particle
system is then evaluated (numerically) as $E=(\int\psi^{*}H\psi dx_{1}\cdots dx_{N})/(\int\psi^{*}\psi dx_{1}\cdots dx_{N})$.
Unfortunately, because of computational limitations, the integration
restricts this study to systems with only $N=3$ particles. For comparison,
the ground state energy obtained with a configuration interaction
calculation ({}``exact diagonalization'') is used as reference.

\begin{figure}
\begin{centering}
\includegraphics[width=1\textwidth]{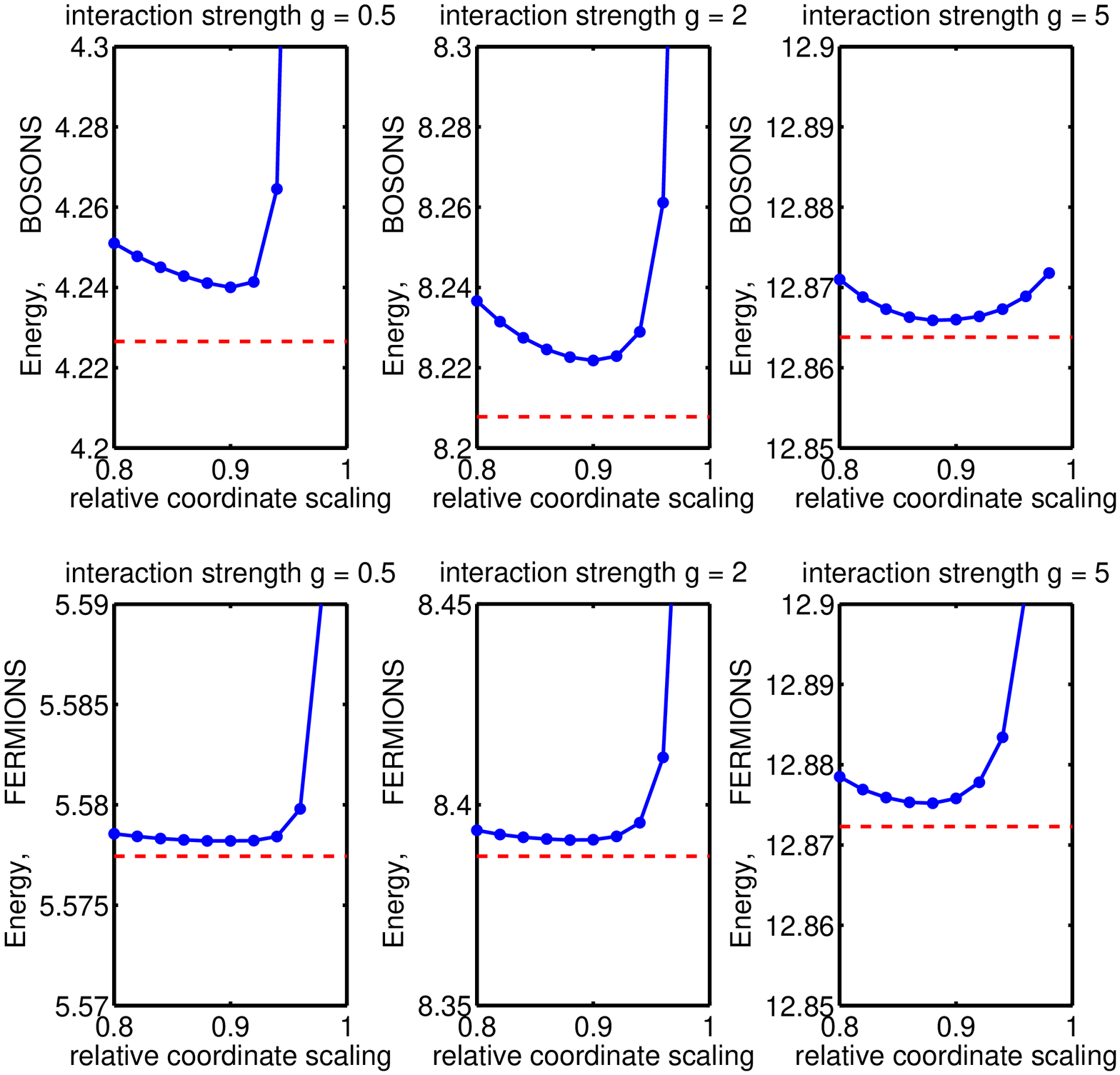}
\par\end{centering}

\caption{\label{fig:N3}Results for $N=3$ particles. The blue dots show the
energy of the trial wavefunction as a function of the scaling parameter
$\alpha$. The dotted red lines show the energy from a configuration
interaction calculation, included as reference. It should be noted
that all values shown are still numerical approximations and only
upper bounds. The reference energies are all converged in the calculation,
and the errors estimated to be very small compared to the scale of
the graphs. The energies obtained with the trial wavefunction also
suffer from similar truncation errors, due to the linear expansion
used for $\phi(x)$.}
\end{figure}

For $N=3$ particles, figure \ref{fig:N3} shows how the energy depends
on the scaling parameter $\alpha$. An electrostatic Coulomb interaction
with prefactor $g$ is used, $V(x)=g/\vert x\vert$, regularized by
assuming that the system is quasi-one-dimensional, with a tight harmonic
confinement in the transverse directions (the transverse oscillator
length is set to $10\%$ of that in the axial direction). The figure
shows results for different interaction strengths, and for both bosons
and (spin-polarized) fermions. In all cases, the energy has a minimum
around $\alpha=0.9$. Based on the data at hand it is of course not
possible to draw any general conclusions about e.g. the $N$-dependence
in $\alpha$.

For $g=0.5$, the system is almost in a perturbative regime, where
the ground state wavefunction is very similar to that of non-interacting
particles. But for $g=5$, the repulsion is so strong that the particles
localize at individual positions (so called Wigner localization),
demonstrated by those energies being very similar for both bosons
and fermions. In either case, a wavefunction with the structure of
equation \ref{eq:trial-wavefunction-N} apparently can give a good
approximation of the ground state. It is worth mentioning that the
first excited state is one energy unit above the ground state, corresponding
to an excited center-of-mass motion of the entire system. Since the
energy of the trial wavefunction is close to the actual ground state,
only a small fraction of the wavefunction can consist of states that
are orthogonal to this.

To summarize, the results imply a simple recipe for how to get an
approximative few-body wavefunction, given the two-body solution.
While not intented as a practical approach for calculations, it does
offer some intuitive insight in the structural properties of the ground
state.
\begin{acknowledgments}
This study was supported by the Swedish Research Council, and the
Nanometer Structure Consortium at Lund University.\end{acknowledgments}

\end{document}